\input harvmac

\def\cst {{\rm const.}}
\def \s {\sigma}

\def \ha {\half}
\def \ov {\over}

\def \lr { \lref}

\def\vp {\varphi}
\def\del {\partial }

\def\l{\lambda}

\gdef \jnl#1, #2, #3, 1#4#5#6{ { #1~}{ #2} (1#4#5#6) #3}

\def\np {  Nucl. Phys. }
\def \pl { Phys. Lett. }

\def \ijmp {Int. J. Mod. Phys. }
\def \cmp {Commun. Math. Phys. }

\lr \berg { E. Bergshoeff,  E. Sezgin, and P.K. Townsend,
\pl {\bf B189} (1987) 75;
 Ann. Phys. {\bf 185} (1988) 330.} 

\lr\dewit { B. De Wit, M. L\" uscher, and H. Nicolai, 
Nucl. Phys. {\bf B320} (1989) 135.}

\lr\dewitt { B. De Wit, J. Hoppe, and H. Nicolai, 
Nucl. Phys. {\bf B305} [FS 23] (1988) 545.}

\lr\kaku{ M. Kaku, 
hep-th/9606057; hep-th/9607111.}
\lr\bars{ I.  Bars, Nucl. Phys. {\bf B343} (1990) 398.}

\lr\town{  P.K. Townsend, Phys. Lett. {\bf B350} (1995) 184.}

\lr\duste{M.J. Duff and K.S. Stelle, \pl {\bf B253} (1991) 113.}

\lr \bps {I. Bars, C.N. Pope and E. Sezgin, Phys. Lett. {\bf 198} (1987) 455.}

\lr\hora {  P. Horava and E. Witten, Nucl. Phys. {\bf B460} (1996) 506.}

\lr\simo {B. Simon, Ann. Phys. {\bf 146} (1983) 209.}

\lr\simon {B. Simon, J. Func. Anal. {\bf 53} (1983) 84.}

\lr\hoppe{J. Hoppe, \ijmp {\bf A4} (1980) 5235.} 

\lr\toro{D. Fairlie, P. Fletcher, and C. Zachos, \pl {\bf B218} (1989) 203;

J. Hoppe, \ijmp {\bf A4} (1989) 5235;

B. de Wit, U. Marquard, and H. Nicolai, \cmp {\bf 128} (1990) 39.}

\lr\thooft{G. 't Hooft, \cmp {\bf 81} (1981) 267.}

\lr\bars{ I. Bars, in  Proceedings of Norman 1990, 
{\it Beyond the standard Model} (1990).}

\lr\otros{ A. Strominger, hep-th/9512059;
E. Witten, \np {\bf B463} (1996) 383;
K. Becker and M. Becker, \np {\bf B472} (1996) 221;
R. Khuri, hep-th/9602153;
L. Cooper and I. Kogan, \pl {\bf B383} (1996) 271;
F. Aldabe, hep-th/9603183;
R. Dijkgraaf, E. Verlinde, and H. Verlinde, hep-th/9604068;
 A. Jevicki, hep-th/9607187.
}
\lr\aha {O. Aharony, J. Sonnenschein, and S. Yankielowicz, hep-th/9603009. }

\lr\schw{ J. Schwarz, hep-th/9607201.}

\lr\toww {P. Townsend, \pl {\bf B373} (1996) 68; 
E. Bergshoeff and M. de Roo, hep-th/9603123 ; M. O' Loughlin, hep-th/9601179;
M. Green, C. Hull, and P. Townsend, hep-th/9604119.}

\lr\duff{ M. Duff, T. Inami , C. Pope, E. Sezgin, and K. Stelle,
\np {\bf B297} (1988) 515.}

\baselineskip8pt
\Title{\vbox
{\baselineskip 6pt{\hbox{CERN-TH/96-251}} {\hbox{hep-th/9609043}} {\hbox{
   }}} }
{\vbox{\centerline {Stability of the quantum supermembrane}
\centerline { in a manifold with boundary}
}}
 
\vskip -20 true pt

\centerline  { J.G. Russo
 }

 \smallskip \bigskip
 
\centerline{\it  Theory Division, CERN}
\smallskip

\centerline{\it  CH-1211  Geneva 23, Switzerland}

\bigskip\bigskip\bigskip
\centerline {\bf Abstract}
\bigskip

We point out an effect which may stabilize
a supersymmetric membrane moving on a manifold with  boundary,
and lead to a light-cone Hamiltonian with a discrete
spectrum of eigenvalues. 
The analysis is carried out explicitly for a closed supermembrane in the regularized $SU(N)$ matrix model version.

\medskip
\baselineskip8pt
\noindent

\Date {September, 1996}

\noblackbox
\baselineskip 16pt plus 2pt minus 2pt

\vfill\eject
 
\noindent 1. In ref.~\dewit\ it was shown that the light-cone 
Hamiltonian of the 
eleven dimensional supermembrane theory \berg\ 
has a
continuous spectrum. The proof was given for  membranes moving in
eleven-dimensional Minkowski space-time.
It is important to investigate whether this instability persists
if the manifold has a boundary.
In particular,
 recent results in string dualities \hora\ indicate that  the strong coupling limit
of ten-dimensional heterotic string theory is described by an
 eleven-dimensional theory 
compactified on $S^1/{\bf Z}_2= I$ ($I$ is the unit interval). 
It is plausible that the eleven-dimensional supermembrane theory 
on the orbifold ${\bf R}^{10}\times I$ may be of  
relevance to this theory.
Another view is to regard the membrane states just as solitonic objects which need not to be quantized in order to have a description of gravitons 
(see e.g. \refs {\hora \duste - \town }). 
In either case,
it would be desirable to identify  possible stable membrane states, which are 
not protected topologically. 

The relevance of the present results to  M-theory on
$S^1/{\bf Z}_2$ is nevertheless unclear 
(for a  review on M-theory, see ref.~\schw ). 
In particular,  
we do not know whether the  boundary conditions  for the membrane wave
function that will be used here can be consistently implemented
in this theory.
In conventional quantum field theory, a boundary in space-time requires
imposing (e.g. Dirichlet or Neumann) boundary conditions on the fields,
to prevent momentum and quantum  information from leaking out 
from the physical space. In particular, a  wave function with 
quantum mechanical probabilistic interpretation 
must vanish at the boundary.
This is the  assumption that will be made here for the
membrane wave function.
In the orbifold $S^1/{\bf Z}_2$, however, the only restriction is that wave functions are ${\bf Z}_2$-invariant, modulo a suitable action on the internal
quantum numbers.
But they are generally not required to vanish at the fixed points; 
one is essentially dealing with a compact space, where the states 
which are not invariant under
the action of the discrete group have been projected out from the Hilbert space.
Thus the only direct implication of the present results to a supermembrane
 moving on
${\bf R}^{10}\times S^1/{\bf Z}_2 $ is perhaps the observation that  its instabilities  are all caused by modes whose associated wave functions do not vanish at the fixed points. In other words, the spectrum of the supermembrane
is discrete upon the restriction to the Hilbert space of membrane states with
nodes at the fixed points.

At long wavelengths,  the dynamics
of the supersymmetric membrane is dictated by the action of \berg\  
(recent discussions on different aspects of
membranes and five-branes can  be found 
in refs.~\refs {\otros ,\aha  }).
Classically,  the presence of a boundary in the space
may appear to have no influence in
solving the instability problem, since the wave function could leak out
to infinity along one of the remaining Minkowski spatial directions.
Fortunately, this is not what happens in the quantum theory: we shall see that 
the boundary modifies the asymptotic zero-point energy of oscillators 
which are
transverse to the
potentially dangerous  direction, in such a way   the resulting motion
of the membrane modes are confined (analogue to the Casimir effect which
stabilizes the bosonic membrane).

We start with the example which in
ref.~\dewit\ was used to illustrate the instability of the supermembrane.
The Hamiltonian is
\eqn\ham{
H= \ha  \left(
\matrix {
- \Delta + x ^ 2 y ^ 2 & x + i y  \cr  
x - i y & - \Delta + x ^ 2 y ^ 2 \cr } \right) \ .
}
The quantum mechanical system we want to consider is governed by the Hamiltonian
\ham\ and, in addition, we put an infinite barrier at $xy=0$, 
the accessible space being defined by  $xy\geq 0$. 
This breaks supersymmetry explicitly. (The  breaking may also be
regarded as ``spontaneous'', if the condition $\psi (0)=0$ is interpreted as a physical restriction on the Hilbert space. It is also possible to define
the system \ham , including the infinite potential barrier, as a limit of a supersymmetric quantum system, i.e. with $H= \ha \{ Q, Q^\dagger \} $~).
Following \dewit , we consider the  wave packet
\eqn\pak{
\psi_t(x,y)=\chi (x-t) \vp_0(x,y)\xi _{\rm F}\ ,\ \ \  \ 
\xi_{\rm F}={1\ov \sqrt{2}}
\left( \matrix {1  \cr -1 \cr } 
\right)\ ,
}
which is designed so as the wave packet  may escape along the potential valley at $y=0$.
Here $\chi (x)$ is a smooth function with compact support, and
the spinor $\xi_{\rm F}$ is chosen to give a maximal negative contribution to the energy of the wave packet, i.e.
$$
\xi_{\rm F}^{\rm T} H\xi_{\rm F}=H_{\rm B} -\ha x\ ,\ \ \ \ 
H_{B}=- \ha \Delta + \ha  x ^ 2 y ^ 2   \ .
$$
We would like to study the motion of $\chi $ in the ground state
of $\vp_0 (x,y)$. $\vp_0 (x,y)$ represents oscillations of the $y$-coordinate
at fixed $x$
about the bottom of the potential valley:
\eqn\ppp{
H_2\vp_0 = E_0 \vp_0\ ,\ \ \ \ 
H_2\equiv -\ha {\del^2\ov \del y^2} +\ha x^2 y^2    \ , 
}
where $H_2$ is viewed as an operator on ${\cal H}_y=L^2({\bf R}, dy)$~.
This is  the equation of the harmonic oscillator. In the absence of a 
boundary, the ground state energy is  $E_0=\ha |x|$, which would
just cancel against the fermionic contribution, and the motion for $\chi $ would be unbounded. Because of the wall, $\vp_0 $ must satisfy the boundary condition
$\vp_0 (y=0)=0$. The wave-function of the ground state is
\eqn\gro{
\vp_0(x,y)= {\sqrt{2}\ov \pi^{1/4} } |x|^{3/4} y\exp(-\ha |x| y^2) \ ,
}
with a zero-point energy
$E_0={3\ov 2} | x|$.
Consequently, we have for $\chi $
\eqn\stab{
\lim _{t\to \infty } (\psi _t, H\psi_t )=
\int dx \chi ^*(x)\big( -\ha \del_x^2+ |x| \big)
\chi (x) \ ,
}
which implies that the motion is bounded, and that the spectrum
of $H$ is discrete (from eq. \stab\ it can be explictly proven that 
${\rm Tr}\ e^{-tH}< \infty $,  e.g. using the ``sliced
bread inequalities" \simon ).

This mechanism relies on the uncertainty principle; the zero-point
energy 
increases as the classically accessible region for the particle is squeezed. Clearly, a similar effect holds for more general boundaries,
for example, a boundary of the form $-1\leq y \leq 1 $
would also increase the zero-point energy of $\vp_ 0$.

\bigskip
\def\T{{\rm T}}
\noindent 2. Let us consider a quantum system governed by the 
following Hamiltonian (analogue to  eq. \ppp )
\eqn\pppp{
H_2 = -\ha {\del\ov \del y_i}{\del\ov \del y^i}
 +\ha (z^\T z)_{ij} y^iy^j\ ,\ \ \ \ i,j=1,...,n  \ ,
}
where $z$ is an $n\times n$ antisymmetric  real matrix.
For $n>2$ the matrix $z^\T z$ will in general not be diagonal.
By a suitable $SO(n)$ rotation $M$, $u_i=M_{ij} y^j$, we can take the Hamiltonian to the form
\eqn\dddd{
H_2 = -\ha {\del\ov \del u_i}{\del\ov \del u^i}
 +\ha w_i^2 u_i u^i\ .
}
First, let us investigate the effect of a wall at $u_i=0 , \ i=1,...,n$, on the 
zero-point energy of this system. The eigenvalue equation is decoupled,
and the result is similar to the example of the previous section,
\eqn\eenn{
E_0={3\ov 2} \sum_i w_i\ .
}
Next, suppose that we change the orientation of the wall, which we now
place at $y_i=0$. Finding the ground state energy of this system
is a complicated problem, and perhaps not solvable by analytic methods.
But there is a generic feature which can be  stated as follows.
\smallskip
\noindent {\it Theorem 1.} Let a quantum mechanical system be described by the  Hamiltonian \dddd\
with a potential barrier  $y_i\geq 0,\ i=1,...,n $, with $y_i=M'_{ij}u^j$.
Then, for all $M'\in SO(n)$ and all $w_i\in {\bf R}$, $w_i\neq 0$, its ground state energy 
is greater than the ground state energy of the similar system without walls.

\smallskip
A simple proof is as follows. Locally, the eigenvalue equation
for both systems is the same, the difference being that in one of them
there is the additional requirement that the wave function vanishes on the boundary. From the standpoint of the system without walls, these
wave functions are just particular solutions with nodes at $y_i=0$.
By the oscillation theorem of quantum mechanics, the expectation value
of the energy on any of these states (in particular, on the normal state
of the system with the walls) must be greater than the ground-state 
energy.
Intuitively, the presence of the wall, irrespective of the orientation,
reduces the classically allowed region; by virtue of the
uncertainty relation, the ground state energy must increase.\foot{
For certain configurations of $w_i$, the statement is
perhaps rather obvious.
If $w_1$ is the lowest frequency, we write
$H_2 = H_2'+ H' ,\   H_2'=-\ha {\del\ov \del u_i}{\del\ov \del u^i}
 +\ha w_1^2 u_i u^i ,\  
H'= \ha \sum_{i=2}^n (w^2_i-w_1^2) u_i^2 . $
Since $H'>0$, the ground-state energy of $H_2$ satisfies $E_0>{3\ov 2}n
w_1$, which is sufficient to demonstrate the theorem for those  frequencies satisfying
${3\ov 2}n w_1> \ha \sum_i w_i,\    {\rm or}   \ 
 w_1> (3n-1)^{-1} \sum_{i=2}^n w_i  $. For other frequencies $H'$
cannot be ignored, since $(w_i^2-w^2)$  will not be small for all $i$.} 

The basic property that will be used in the next section is contained
in the following theorem, which is a generalization of the former
to the case when the boundary is imposed in terms of a Fourier-like transform.
\smallskip
\noindent {\it Theorem 2.} Same as theorem 1, but now the potential barrier is
$X(\s )\geq 0 \ ,\ \ X(\s )=\sum _i u_i f_i (\s )$, where
the $f_i (\s )$ represent a  set of orthonormal functions on some space.
\smallskip
Consider the function $X(\s )\equiv 0 $. Because of orthonormality,
this corresponds to the choice $u_i =0$ for all $i$.
Thus, at the point $u_i=0, i=1,2...$, the eigenfunctions of the system with walls must have a node. On the other hand, the system without walls is a collection of decoupled harmonic oscillators. The ground state is just the direct product of the ground states of individual oscillators,
and it does not have any node. Therefore its energy must be lower, as stated above.

\bigskip
\noindent 3.
By expanding the coordinates in a complete orthornormal basis of
functions $Y^A(\s )$ on the membrane, ${\bf X}(\s )=\sum_{A} {\bf X}^A Y_A(\s )$,
and similarly for the fermionic variables and the momenta, the 
 light-cone gauge  Hamiltonian takes the form \dewitt
\eqn\lcgh{
H = { 1\over 2 }  P ^A_\mu P_{\mu A} + {1\ov 4} f_{ABE} f_{CD}^E X^A_\mu X^B_\nu X^C_\mu X^D_\nu 
-{1\ov 2} i f_{ABC} X^A_\mu \theta ^B \gamma ^\mu \theta^c\ ,
}
where $\mu, \nu = 1,2,...,9$ and $f_{ABC}$ are the structure constants of
the group of area-preserving diffeomorphisms of the parameter manifold.
Here  $\theta ^A_\alpha $, with $\alpha =1,...,16$, are real $SO(9)$ spinors.
The classical instabilities occur along the Cartan directions, where the potential vanishes. For concreteness, we  will consider the case of
spherical or toroidal membranes, where the group can be regarded as $SU(N)$,
with $N\to \infty $. The generalization to other compact Lie groups should be straightforward.
It is convenient to split off the coordinates $X^A_\mu $ into the form
$X^A_\mu \to (Z^i_\mu, Y^I_\mu)$
where indices $i,j,k=1,..., N-1$ correspond to the Cartan directions,  and 
$I, J= N,..., N^2-1$ label the remaining directions. Upon gauge fixing,
where $Y^I_9 $ is removed,
only a residual invariance under the Cartan subgroup remains,
and the Hamiltonian takes the form \dewit
\eqn\hamil{
H = H _ 1 + H _ 2 + H _ 3 +  H _ 4 \ ,
}
\eqn\hhdd{
H _ 1  = - { 1 \over 2 }
\left ( { \partial \over \partial Z^k }\right) ^ 2
-{ 1 \over 2 }
\left ( {\partial \over \partial Z _ a^k } \right ) ^2\ ,
\ \ \ H_2=-\ha \left( \del\ov\del Y_a^I \right)^2
+\ha  (z^\T z)_{IJ} Y^I_a Y^J_a \  ,
}
\eqn\hhjj{
H _ 3 = - \ha i
 \theta ^I \big( z_{IJ} \gamma _ 9 + 
z ^ a _{IJ} \gamma _ a  \big) \theta ^J\ ,
}
$$
z_{IJ}^a=Z^{ak} f_{kIJ}\ ,\ \ \  Z^k=Z_9^k\ ,
\ \  \ a=1,...,8   \ .
$$
$H _ 4 $  is not important in the analysis
of  stability, and here it will not be considered (as  in \dewit ,  all terms in $H_4$ vanish in the asymptotic region).
The eigenvalues of $z_{IJ}$
 are  the (non-vanishing) roots of the Lie algebra of $SU(N)$,
and can be expressed in terms of the eigenvalues $i\l _m ,\ i\l _m ^a ,\ 
\l_m \in {\bf R},\ m=1,...,N $ of $Z$ and $Z_a $. The following relations will be useful:
\eqn\rrrr{
\det z=\prod _{m<n }^N (\l _m - \l _n )^2 =\det \Omega\ ,\ 
\ \ \ {\rm tr}\ \Omega=2\sum _{m<n}^N (\l _m-\l_n)\ ,
\ \ \ \ \Omega\equiv \sqrt{z^\T z} \ .
}
 
Now we consider the  wave-packet  which in ${\bf R}^{11}$ causes a instability (representing a mode escaping along the
potential valley). This has the form
\eqn\pakk{
\psi _t (Z,Z_a, Y^I_a) = \chi (Z - t V, Z _ a ) 
\phi _ 0 ( Z , Y _ a^I ) \xi _{\rm F} ( Z , Z _ a ) \ ,
}
\eqn\pakkk{
V_{mn}=i\big[ \ha (N+1)-m \big] \delta _{mn}\ .
}
The fermionic variable has the ground state energy 
\eqn\fff{
H _ 3 \xi = - 8 \sum _ { m<n } \omega _ { mn }\xi \ ,
}
where
\eqn\www{
\omega _ { mn }  =   \sqrt { \lambda _ { mn } ^ 2 
+ ( \lambda _ { mn }^ { a} ) ^  2  }  \ ,\ \ \ \l_{mn}=\l_m - \l_n \ .
}
As far as the fermionic ground state energy is concerned, the presence of a boundary, e.g. of the type  $Y^I_{8}\geq 0$, is  irrelevant
($H_3$ does not even depend on $Y_a^I$).
The derivation of \fff\ is identical to the derivation of
 ref. {\dewit}, so it will not be reproduced here.

Now we determine the zero-point energy of $H_2$ (see eq. \hhdd ).
Denote by  $U_a^I$ the basis in which
$H_2$ is diagonal, where $U_a^I$
is related to the $Y_a^I$ by an orthogonal transformation, 
$$
 U_a^I=  M_{IJ} Y_a^J \ .
$$
Let us first investigate the effect of a ``boundary"
imposed in terms of $U_{8}^I$, $U_{8}^I\geq 0$ for all $I $.
In this case the eigenstates of $H_2$ can be explicitly constructed, 
being a straightforward extension of the quantum mechanical systems  
considered in the previous sections.
The $U_a$ with $a=1,...,7$ will give a contribution to the zero-point energy
proportional to ${7\ov 2} {\rm tr}\ \Omega $,
whereas the contribution of $U_8$ will be proportional to 
${3\ov 2} {\rm tr}\ \Omega $. Explicitly,
\eqn\hhhh{
H _ 2 \phi  _ 0 = 5 \big( {\rm tr}\ \Omega \big)\ \phi _ 0 \ ,
}
where $\phi_0 $ is given by (cf. eq. \gro )
\eqn\pphh{
\phi _ 0 (Z, Y_a^I )= \cst ( {\rm det } \, \Omega ) ^ {5/2} 
\left ( \prod_{I} U_8 ^I \right) \
\exp \left ( - { 1 \over 2 } \Omega _{IJ}  Y _ a ^ I Y_a^J   \right)
 \ .
}
The potential for $\chi $ arises from the contribution of
$H _ 2$ and $H _3$ to the total energy:
\eqn\tott{
( H _ 2 + H _ 3 ) \psi _t = 2
\sum _ { m< n}^N \left ( 
5|\lambda _ { mn }| - 4\omega _ { mn } \right ) \psi _t   \ .
}
For large $t$ (see eqs. \pakk , \pakkk ),
\eqn\llll{
\lambda_ { mn }=(n-m) t +  O(1)\ ,\ \ 
}
and $\omega _{mn} \to |\l_{mn }|$. As a result we get
$$
\lim_{t\to \infty} || (H_2+H_3)\psi _t ||= {\rm tr}\ \Omega=
\left( 2\sum_{m<n}^N (n-m) \right) t \ .
$$
Thus  $\chi $ cannot move off to infinity; the
system can only execute a finite motion in the $Z$ direction.
The theorem 1 states that the same conclusion applies
when the boundary is imposed in terms of the  
 coordinate $Y_8^I$, i.e. $Y_8^I\geq 0$, 
related by an orthogonal rotation to the $U^I_8$.
By  theorem 2, the
 motion of $\chi $  will also be finite when the boundary is imposed in terms of the physical
 coordinate $X_8(\s )$, i.e. $X_8(\s )\geq 0 $ for all $\s $.

Nevertheless, for this system with a single wall, the spectrum will be continuous
in virtue of the fact that there is
still a direction along which a mode can leak out
to infinity. This is the direction orthogonal to the wall, say,
orthogonal to the hyperplane $Y_8=0$.
Indeed, for large values of $Z_8$, 
the boundary has no effect in the oscillators transverse to this direction. To see this explicitly,
we gauge away $Y_8$ (instead of $Y_9 $) and consider a similar wave packet with
$\chi (Z_8-tV )$. Then $H_2$ only involves  coordinates which are unbounded,
so that its ground state energy is $4 \ {\rm tr}\  \Omega $, which
in the asymptotic region
cancels against the fermionic contribution.
To  eliminate the possibility of infinite motion in this direction, we generalize the above discussion by adding an extra wall. In particular,  systems with either an extra boundary
component, such as $0\leq Y_8\leq 1 $, or a boundary in another
direction, e.g. $Y_8, Y_7\geq 0$, do not have that possibility
of leakage.
The motion in all Cartan directions is finite and the spectrum of 
eigenvalues must therefore be discrete.

It may seem counterintuitive that the  inclusion of a 
boundary in a single dimension can stabilize the supermembrane.
What happens is that all the coordinates couple to the same
matrix $ (z^\T z)_{IJ}$. A change of the zero-point energy of
a single coordinate modifies the total zero-point energy of $H_2 $ by a numerical
factor; it no longer cancels against the fermionic contribution and a confining potential for $\chi $ is generated.
 The discussion can be formally generalized to the 
 continuum, by using a framework recently introduced in ref.~\kaku .

\bigskip

4. So far our discussion has only included spherical and toroidal membranes.
A similar analysis applies for open membranes of cylindrical topology.
Indeed, the basis functions are essentially equivalent to those of the
torus \toro , $Y_{\rm K}(\s )= e^{ik_1\s_1 +ik_2\s_2} $, $K=(k_1,k_2)$, 
$\s_{1,2}\in [0,2\pi )$.
By making use of 't Hooft's twist matrices \thooft , it is possible to 
construct exactly $N^2-1$ traceless independent matrices $T_K$, satisfying an algebra
which approaches the Lie algebra of $SU(N)$, as $N\to \infty $.

 An open supermembrane with boundary components living on the
boundary of the manifold ${\bf R}^{10} \times I$,
perhaps of relevance to the strong coupling limit of heterotic string theory, is likely to have a discrete spectrum as well (in the restricted Hilbert space
of membrane wave functions which vanish at the space-time boundary).
The only difference lies on the dynamics of the membrane boundary, 
represented by closed strings propagating in ${\bf R}^{10}$,
but this dynamics should not spoil stability.

Another question concerns the dependence of the eigenvalue spectrum of the Hamiltonian  on the membrane topology.
The analysis using the regularized Hamiltonian does depend on the properties
of the specific Lie group, such as  roots and structure constants.
However, it is believed  that
the Poisson algebra of functions on a manifold that is a regular coadjoint orbit of some group $G$ can always be approximated by $SU(N)$, with possible restrictions to some subalgebra (see e.g. refs.~\refs{\toro ,\bars }).
Some dependence of the spectrum on the topology is expected,
given that different membrane topologies should 
in fact represent inequivalent quantum states. 
In the eleven dimensional theory on ${\bf R}^{11}$ there is no 
coupling parameter that can suppress higher topologies.
On the other hand, in the compactified theory,
by combining the membrane tension  $T_3$ and  the size $R_{11}$ of the eleven dimension, it is possible to define an adimensional  coupling parameter 
$g^2=T_3 R_{11}^3$ (which in the standard correspondence with heterotic superstring theory  
is in turn related to the dilaton field \hora). 
%
%
The way  the D-2-brane is  quantized in 
type IIA superstring theory suggests that
a consistent quantization of the eleven dimensional membrane  
may  require including states associated with higher topologies.
Adding small handles should not considerably increase
the mass, since the cost of energy is proportional to the 
membrane area.
Just as would be the case for a Ramond-Ramond soliton 
of type IIA string theory, for  $g^2=O(1)$
the low-energy excitations of the membrane should be generically 
constituted of both oscillation modes and tiny handles.
In this case a more adequate formalism may rather involve
a suitable quantization of the world-volume field theory  \toww .
While the D-brane picture is not justified in extrapolating from
from weak  to strong string coupling, where the eleven dimension emerges,
it is indicative in discriminating the relevant
degrees of freedom in the various limits of the product $T_3 R_{11}^3$
(see also ref.~\aha ).

\bigskip

5. 
A priori, possible  unbroken supersymmetries do no 
imply a relation between the
zero-point energies  of  $H_2$ and $H_3$. 
Note that what is calculated here is just the zero-point energy 
of $H_2+H_3$ for the particular  decomposition of the trial 
wave function \pakk\  
(in the quantum subsystem ${\cal H}\equiv H_2+H_3$, with $z_{IJ}$ fixed, 
the presence of a boundary breaks supersymmetry).
Although our analysis does not therefore exclude
the existence of  a set of normalizable 
eigenvectors of $H$ with zero eigenvalue
(representing a massless multiplet in the
 supermembrane spectrum), there are several reasons to expect that
the ground state in the stabilized theory will be massive.
As in the $D=10$ superstring theory, in supermembrane theory
the existence of a massless multiplet seems to be ascribed
to a complete cancellation of the bosonic and fermionic
ground state energies \duff .
In the ${\bf R}^{10}\times S^1/{\bf Z}_2$ orbifold compactification, half of the supersymmetries remain, but in the present context they 
could be 
broken because of the boundary conditions.
Just from the zero-mode structure, 
a contingent unbroken supersymmetry, 
together with the discreteness of the spectrum,
could be used to argue that there is a unique ground state, 
constituted by the eleven dimensional supergravity multiplet \bps .
Unfortunately, the general ground-state wave functional is not known, so
it is not presently possible to determine whether the boundary condition is
eliminating a square-integrable massless state.
(The structure of the equation $Q\Psi =0$ actually suggests
that there cannot be any normalizable  massless state in 
supermembrane theory \dewitt , but the problem is  still open).
For compactifications in more complicated topologies, say $T^2/{\bf Z}_2$,
the generic situation is that there is no solution to a differential equation that is everywhere non-vanishing. From this point of view, demanding
the wave functional to have nodes
at the fixed points seems less restrictive.
Clearly, further work is needed.
In particular, it would be of interest to identify the membrane topologies
which can lead to a non-vanishing Witten index, which seems to be
computable in the regularized low $N$ theory.

\bigskip
The author wishes to thank 
L. Alvarez-Gaum\' e, I. Bars, H. Nicolai, P. Townsend, 
and A. Tseytlin for useful comments.

\listrefs

\vfill\eject
\end